\documentclass[conference]{IEEEtran}
\IEEEoverridecommandlockouts
\usepackage{cite}
\usepackage[comma, numbers]{natbib}
\usepackage{amsmath,amssymb,amsfonts}
\usepackage{algorithm,algpseudocode,float}
\usepackage{lipsum}
\usepackage{graphicx}
\usepackage{subfigure}
\usepackage{multirow}
\usepackage{array}
\usepackage{colortbl}
\usepackage{tabularx}
\usepackage{textcomp}
\usepackage{xcolor}
\def\BibTeX{{\rm B\kern-.05em{\sc i\kern-.025em b}\kern-.08em
    T\kern-.1667em\lower.7ex\hbox{E}\kern-.125emX}}

\usepackage[left=0.673in,right=0.673in,top=0.751in,bottom=1.05in]{geometry}

\makeatletter
\newenvironment{breakablealgorithm}
  {
   \begin{center}
     \refstepcounter{algorithm}
     \hrule height.8pt depth0pt \kern2pt
     \renewcommand{\caption}[2][\relax]{
       {\raggedright\textbf{\ALG@name~\thealgorithm} ##2\par}%
       \ifx\relax##1\relax 
         \addcontentsline{loa}{algorithm}{\protect\numberline{\thealgorithm}##2}%
       \else 
         \addcontentsline{loa}{algorithm}{\protect\numberline{\thealgorithm}##1}%
       \fi
       \kern2pt\hrule\kern2pt
     }
  }{
     \kern2pt\hrule\relax
   \end{center}
  }
\makeatother
    
\begin{document}

\title{Selecting Relay Nodes Based on Evaluation Results to Enhance P2P Broadcasting Efficiency}

\author{
	\IEEEauthorblockN{Chunlin Huang} 
	\IEEEauthorblockA{Chengdu University of Information Technology, Chengdu, China\\trenlinhuang@gmail.com}
} 

\maketitle

\begin{abstract}
The existence of node failures is inevitable in distributed systems, thus many P2P broadcasting networks adopt highly robust Flooding-based broadcast algorithms. High redundancy inevitably leads to high network resource consumption, and it may constrain the data transmission rate of the network. To address excessive network resource consumption, many studies have explored broadcasting mechanisms in structured P2P overlay networks. However, existing DHT-based algorithms cannot assess the quality of neighbors, which is crucial for broadcast efficiency. In this paper, we introduce the Neighbor Evaluation mechanism to select relay nodes based on their evaluated contributions. According to experimental results, the Neighbor Evaluation mechanism has a significant effect on both broadcast latency and coverage rate.
\end{abstract}


\section{Introduction} \label{section:introduction}

In broadcasting networks that use Gossip \cite{leitao2010gossip}, whenever a node receives a new message, it relays the message to a specific number of nodes among its neighbors, excluding the sender. If a node has already received the data, it will simply ignore this message. Gossip can tolerate node crashes to a certain extent and messages can often reach other nodes via the optimal path. In a network with N nodes, each node except the sender only needs to receive each message once, which means that there are only N-1 effective transmissions for each broadcast. Although high redundancy of the Gossip ensures broadcast coverage rate well, the amount of transmitted messages is obviously much greater than N-1, which means that a lot of additional network resources are consumed. Furthermore, compared to Flooding that relays messages to all neighbors, even in the absence of any faulty or malicious nodes in the network, Gossip cannot guarantee the propagation of messages to every single node.

\begin{figure}[htbp]
    \centering
    \includegraphics[scale=0.4]{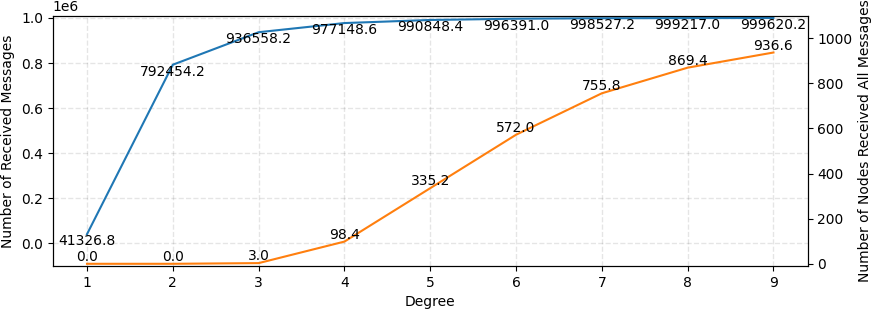}
    \caption{Broadcast coverage of gossip algorithm in a simulated 1000-node network under different redundancy degrees. Each value in the figure represents an average value of five simulation experiment results.}
\end{figure}

Distributed Hash Table (DHT) is a decentralized system that provides routing and lookup services. For the redundancy issue in broadcast in unstructured P2P networks, the broadcast algorithm based on structured DHT can solve it well. Research \cite{el2003efficient} creatively proposed a structured broadcasting algorithm based on DHT, which can transmit information to all nodes in the P2P network with no redundancy. However, as mentioned in \cite{wang2022data}, the loss of any data packet in DHT-based broadcasting network may cause the early termination of propagation paths and potentially result in a considerable number of unreached nodes.

\section{Related Work} \label{section:related-work}


Research \cite{el2003efficient} developed the first P2P broadcasting algorithm in structured network based on Chord \cite{stoica2001chord}. Broadcast algorithms based on Kademlia \cite{maymounkov2002kademlia} have been implemented in \cite{peris2012evaluation, czirkos2013solution}, which use redundancy strategy to some extent to maintain coverage rate when the network experiences disturbances. Kadcast \cite{rohrer2019kadcast}, based on the Kademlia, implements a broadcast algorithm for block propagation in blockchain network. Swift \cite{wang2022data} uses a broadcast method based on an algorithm similar to CAN \cite{ratnasamy2001scalable} to broadcast blocks. It maps nodes into a multi dimensional coordinate space, recursively dividing the entire P2P network to many non-overlapping propagation scopes. However, compared to Flooding-based broadcast algorithms, the weakness of structured network-based broadcast algorithms lies in their reliability. Although redundancy strategies can bring some reliability, they are not sufficient. Therefore, selecting more reliable neighbors as relay nodes could also be a strategy to enhance broadcast efficiency.


\section{Mechanism Design} \label{section:mechanism-design}
\subsection{Observations and Design Principles} \label{observations-and-design-principles}
In a decentralized overlay network, nodes' awareness of the network is derived solely from their data exchanges with a limited number of its neighboring nodes. The information that nodes can obtain from the data exchanges include:
\begin{itemize}
    \item The online status of neighbor nodes. This information can be used to remove or replace neighbor nodes that are offline.
    \item Whether the received message is a redundant message or a new message. It determines whether the message should be ignored or received, verified, and forwarded.
    \item Which neighbor node the received message comes from.
\end{itemize}

These three types of information constitute the basis for evaluating the quality of neighboring nodes.

\subsection{Neighbor Evaluation Mechanism}
\begin{figure}[htbp]
    \centering
    \includegraphics[scale=0.32]{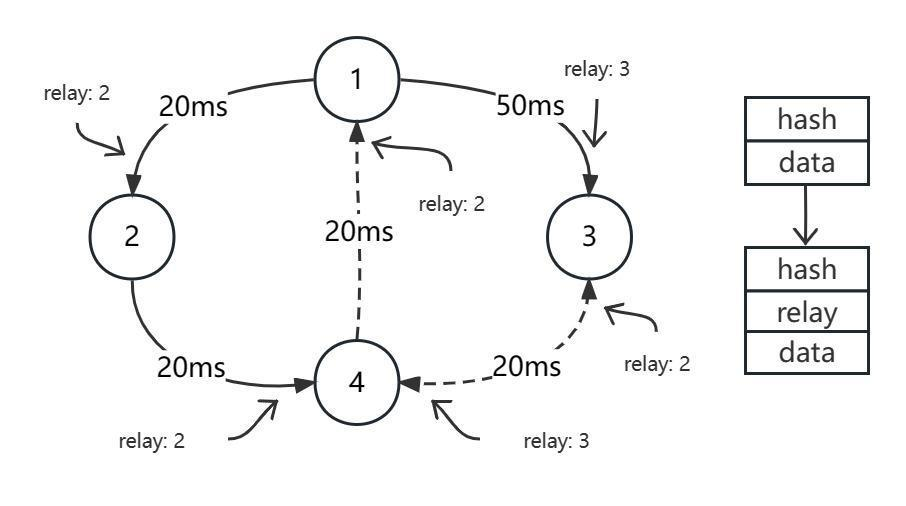}
    \caption{In order to enable the source node to identify which relay node the redundant message comes from, a relay field is added to the message.}
\label{fig:add-relay}
\end{figure}

In the perspective of the source node, when it initiates a broadcast and sends messages to some neighbor nodes, in a connected network, it will receive redundant messages relayed by some other neighbor nodes. We denote the former as relay nodes and the latter as probe nodes. For example, in a four-node network as illustrated in Fig. \ref{fig:add-relay}, after the source node 1 broadcasts a message to relay nodes 2 and 3, it will receive a redundant message from probe node 4. Despite the redundancy inherent in these messages, their absence would render the source node incapable of determining whether the message it disseminated has been forwarded into the network by relay nodes it selected before. Moreover, It can be seen from Fig. \ref{fig:add-relay} that the message relayed from node 4 to node 1 relays from the neighbor node 2, but the source node is not aware of this information because all nodes relay the identical messages. Therefore, we need to modify the message of the broadcast by adding a \textit{relay} field to indicate which neighbor node of the source node the message is relayed from. With this adjustment in place, the quantification of relay nodes' contributions in the broadcasting process becomes feasible.

\begin{figure}[htbp]
    \centering
    \includegraphics[scale=0.32]{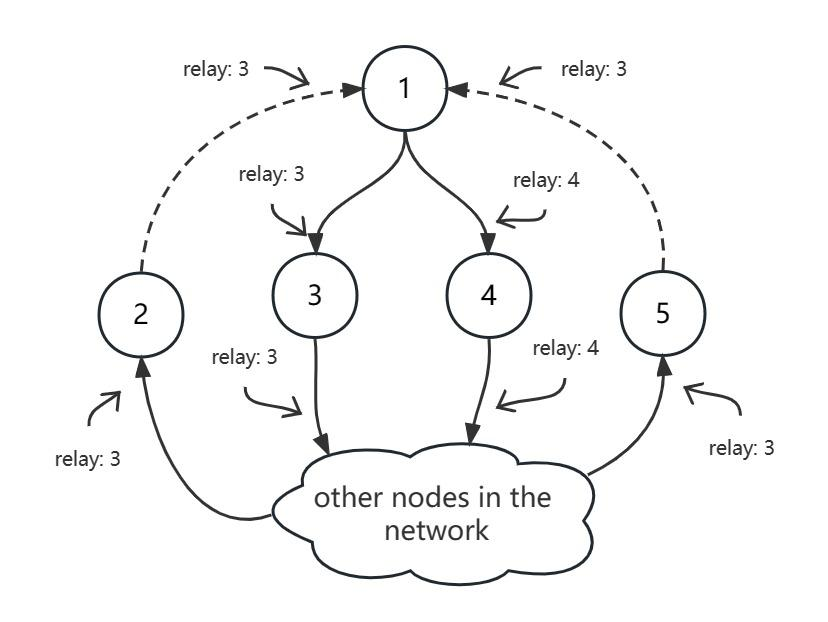}
    \caption{Competition among relay nodes during a broadcast. In the figure, source node 1 sends messages to nodes 3 and 4, and then receives messages relayed by node 3 from nodes 2 and 5.}
\label{fig:broadcast-competition}
\end{figure}

As mentioned earlier, nodes' perception of the network comes only from interactions with a limited number of nodes it connected with. Each broadcast process can be seen as a round of voting by probe nodes for relay nodes. In Fig. \ref{fig:broadcast-competition}, because the message relayed by node 3 was propagated more quickly in the network, it arrived at probe nodes 2 and 5 first and was then relayed to the source node. Since the process of neighbor relationship establishing between nodes in the network can be seen as random, the voting process can better reflect which relay nodes' messages are more quickly broadcasted in the network when there are sufficient neighbor nodes. 

\subsubsection{Quantifying the Contribution of Neighbors}
For the source node, the quality of its neighbor nodes is reflected by the scores in the node information in the routing table.  A neighbor node's score is increased when it behaves in a way that is beneficial to the node's interaction with the rest of the network. Actions that can increase the score of a neighbor node include:
\begin{itemize}
    \item Relaying a new message to the node.
    \item As a probe node, redundantly relaying a message to the source node that was initially broadcasted by the source node itself.
    \item As a relay node, forwarding a message to the network that arrived earlier than other relay nodes at a particular probe node and the message was subsequently relayed to the source node.
\end{itemize}
When a node is detected to be unresponsive, its information in the routing table will be discarded, which means its score will be lost as well.
\subsubsection{Broadcasting in Kademlia with Neighbor Evaluation Mechanism}
\begin{figure}[htbp]
    \centering
    \includegraphics[scale=0.30]{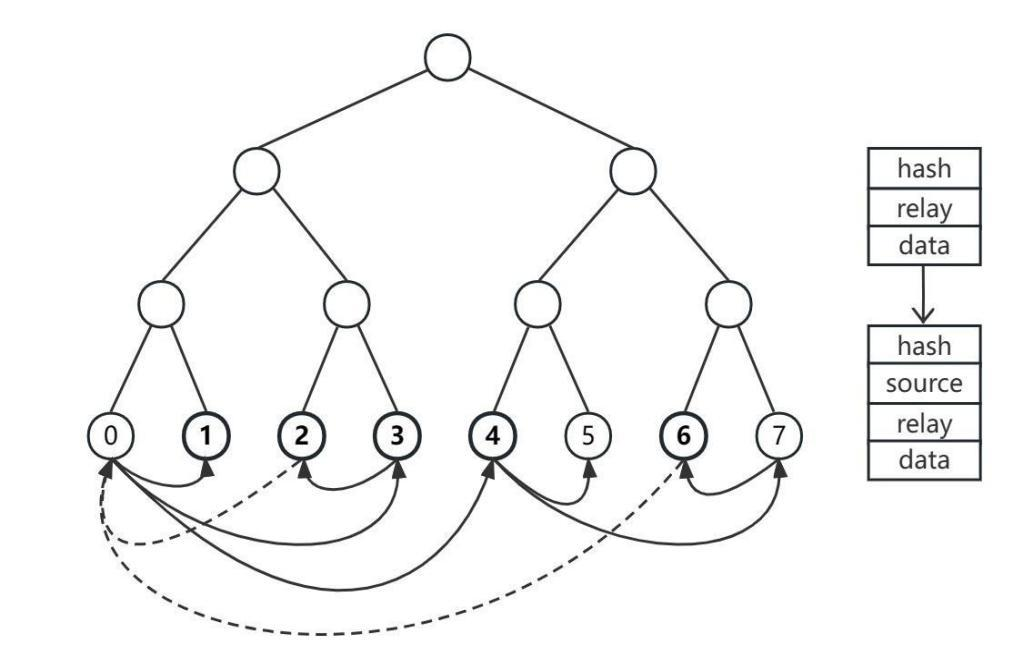}
    \caption{A broadcast in a Kademlia-based network with 8 nodes. Nodes 1, 2, 3, 4, and 6 are neighbors of the source node 0. Among them, nodes 1, 3, and 4 are selected as relay nodes by source node 0, while nodes 2 and 6 are selected as probe nodes  in this round of broadcasting. The solid lines represent the broadcast path of the Kademlia-based broadcast algorithm, and the dashed lines represent the redundant relay introduced in this paper to support the Neighbor Evaluation mechanism.}
\label{fig:add-source}
\end{figure}

As the Kademlia-based broadcast algorithms can be seen as recursively performing a new broadcast in each subtree and no message will be relayed to the source node, it is necessary to modify the message by adding a source node field. And nodes are required to perform a redundant relay to the source node when they receive messages and discover that the source node is their neighbor, as shown in Figure \ref{fig:add-source}. The messages relayed in redundancy are denoted as redundant confirmation packets.

\begin{figure}[htbp]
    \centering
    \includegraphics[scale=0.3]{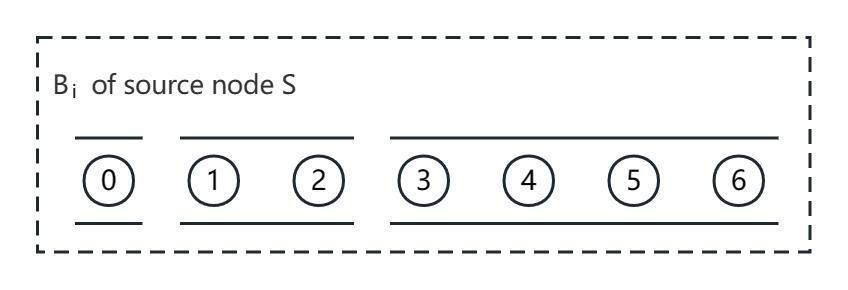}
    \caption{Grouping of nodes in bucket $B_i$ Nodes 3-6 have a weight of 1, nodes 1-2 have a weight of 2, and node 0 has a weight of 4.}
\label{fig:grouping}
\end{figure}

In each bucket of a node's routing table, neighbor nodes are sorted from high scores to low scores, as shown in Fig. \ref{fig:grouping}. And node i is assigned to the $\lfloor log_{2}(i+1) \rfloor$-th group. The size of the bucket should be $2^{n}-1$ (e.g., 7, 15). If there are a total of R groups in the bucket, each node in the r-th group has a weight of $2^{R-r}$. The weight of the node i in the bucket is denoted as $W_i$. The probability that node i is selected as a relay node is the proportion of its weight to the sum of scores of all unselected nodes:
\begin{equation}
    P_i = \frac{W_i}{\sum_{k=1}^{n}W_k - \sum_{k=1}^{m}W'_k}
\label{formula:probability-selection}
\end{equation}
where n represents the number of nodes in a bucket, $W'$ represents the set of nodes that have already been selected, and m denoting the number of nodes in $W'$. This means that in the case where all groups are saturated, each group has an equal probability of being selected. And it ensures that new nodes in a bucket and nodes with low scores still have a chance of being selected, thus preventing the occurrence of starvation caused by a few nodes with extremely high scores. When a node receives a message and relays it, it also follows the same strategy to select its successor nodes. It should be noted that, in a specific broadcast round, only nodes that have not been selected as relay nodes are eligible to serve as probe nodes within each bucket. In other words, a node can assume only one role per broadcast round. And each bucket operates independently without interference from other buckets. The process of initiating a broadcast is shown in Algorithm \ref{alg:new-msg},  and the process of receiving and relaying a message is shown in Algorithm \ref{alg:relay-msg}.

\begin{breakablealgorithm}
    \caption{Initiating a new broadcast}\label{alg:new-msg}
\begin{algorithmic}
    \State Set of known hash of messages: $M$
    \State Set of tickets that probe nodes can vote for relay nodes: $T$
    \State Redundancy factor: $\beta$
    \State Id of this node: $id$
    \State
    \State $message \gets new\_message(data)$
    \State $message.source \gets id$
    \State $message.hash\_computing()$
    \State $M \gets M \cup \{message.hash\}$
    \State for $b$ in Buckets do
    \State \hspace{\algorithmicindent} $relay\_nodes \gets random\_based\_on\_weight(\beta, b)$
    \State \hspace{\algorithmicindent} for all $peer \in relay\_nodes$ do
    \State \hspace{\algorithmicindent}\hspace{\algorithmicindent} $m \gets message$
    \State \hspace{\algorithmicindent}\hspace{\algorithmicindent} $m.relay \gets peer.id$
    \State \hspace{\algorithmicindent}\hspace{\algorithmicindent} $m.sig \gets sign(m)$
    \State \hspace{\algorithmicindent}\hspace{\algorithmicindent} $relay\_message(peer.id, m)$
    \State \hspace{\algorithmicindent} end for
    \State \hspace{\algorithmicindent} $probe\_nodes \gets b - relay\_nodes$
    \State \hspace{\algorithmicindent} for all $peer \in probe\_nodes$ do
    \State \hspace{\algorithmicindent} \hspace{\algorithmicindent} $T \gets T \cup \{new\_ticket(message.hash, peer.id)\}$
    \State \hspace{\algorithmicindent} end for
    \State end for
\end{algorithmic}
\end{breakablealgorithm}

\begin{breakablealgorithm}
    \caption{Message reception and relaying} \label{alg:relay-msg}
\begin{algorithmic}
    \State Received message: $m$
    \State Set of known hash of messages: $M$
    \State Set of tickets probe nodes can vote for relay nodes: $T$
    \State Redundancy factor: $\beta$
    \State Id of this node: $id$
    \State
    \State if $check\_sig(m) \neq true$ then
    \State \hspace{\algorithmicindent} return
    \State end if
    \State if $id = m.source$ then
    \State \hspace{\algorithmicindent} $ticket \gets new\_ticket(m.hash, m.from)$
    \State \hspace{\algorithmicindent} if $ticket \notin T$ then
    \State \hspace{\algorithmicindent} \hspace{\algorithmicindent} return
    \State \hspace{\algorithmicindent} end if
    \State \hspace{\algorithmicindent} $add\_score(m.from)$ 
    \State \hspace{\algorithmicindent} $add\_score(m.relay)$ 
    \State \hspace{\algorithmicindent} $T \gets T - \{ticket\}$
    \State end if
    \State if $m.hash \in M$ then
    \State \hspace{\algorithmicindent} return
    \State end if
    \State $add\_score(m.from)$ 
    \State $M \gets M \cup \{m.hash\}$
    \State if $is\_neighbor(m.source)$ and $m.from \neq m.source$ then
    \State \hspace{\algorithmicindent} $redundant\_relay(m.source, m)$
    \State end if
    \State $h \gets get\_height(id, m.from)$
    \State for $i \gets h$ to $num\_address\_bits$ do
    \State \hspace{\algorithmicindent} $b \gets Buckets[i]$
    \State \hspace{\algorithmicindent} $relay\_nodes \gets random\_based\_on\_weight(\beta, b)$
    \State \hspace{\algorithmicindent} for all $peer \in relay\_nodes$ do
    \State \hspace{\algorithmicindent}\hspace{\algorithmicindent} $relay\_message(peer.id, m)$
    \State \hspace{\algorithmicindent} end for
    \State end for
\end{algorithmic}
\end{breakablealgorithm}

\section{Evaluation} \label{section:evaluation}
In this section, the performance improvement of the NE mechanism on the Kademlia-based broadcast algorithm is evaluated. The size of each broadcast message in the simulation environment is set to 128B, and the size of the redundancy relay packet is set to 20B. The 1000 nodes in the network are randomly assigned to four different regions, each with distinct upstream bandwidths. Bandwidth allocation is shown in Table \ref{tab:net-config-bandwidth}, and the delay between regions is shown in Table \ref{tab:net-config-delay}. In addition, the bucket size k of nodes is set to 15.

\begin{table}[htbp]
    \centering
    \caption{The Proportion of Nodes in Each of the Four Regions, and the Proportion of Nodes with Different Upstream Bandwidths}\label{tab:net-config-bandwidth}
    \begin{tabular*}{\hsize}{@{}@{\extracolsep{\fill}}cc||cc@{}}
    \hline
    Region &Proportion & Upstream Bandwidth (bps) & Proportion  \\
    \hline
    a      & 30\% & 512k & 10\% \\
    \hline
    b      & 10\% & 256k & 60\% \\
    \hline
    c      & 40\% & 128k & 20\% \\
    \hline
    d      & 20\% & 64k & 10\%  \\
    \hline                         
    \end{tabular*}
\end{table}

\begin{table}[htbp]
    \centering
    \caption{Delay between Regions ($\mu$s)}\label{tab:net-config-delay}
    \begin{tabular*}{\hsize}{@{}@{\extracolsep{\fill}}r|c|c|c|c@{}}
    \hline
    & a & b & c & d \\
    \hline
    a & 10,000  & 200,000 & 250,000 & 250,000  \\
    \hline
    b & 200,000 & 3,000   & 100,000 & 100,000  \\
    \hline
    c & 250,000 & 100,000 & 7,000   & 200,000  \\
    \hline
    d & 250,000 & 100,000 & 200,000 & 8,000 \\
    \hline
    \end{tabular*}
\end{table}

\subsection{Simulation Model}

\subsubsection{Performance Measuring Method}
In the following part, the performance of the broadcast algorithms is evaluated by the broadcast latency and coverage rate. The latency of a broadcast is the time from the first node sending the message to the last node receiving the message, and the coverage rate is calculated by dividing the number of received messages in the network by the product of the number of nodes and the number of total rounds. We record a $MessageRec$ record for each broadcast data to calculate its latency and coverage rate. The structure of $MessageRec$ is as follows:
\begin{equation}
MessageRec \{ ReceivedCount, Timestamps \}
\end{equation}

Whenever a broadcast data is initialized, the start time is recorded into $Timestamps$. $ReceivedCount$ is updated each time a message is fetched from the message min-heap. When the $ReceivedCount$ equals the number of network nodes, the time when the last message was received is recorded into $Timestamps$.

\subsection{Broadcast Latency}
\begin{figure}[htbp]
    \centering
    \subfigure{
    \includegraphics[scale=0.4]{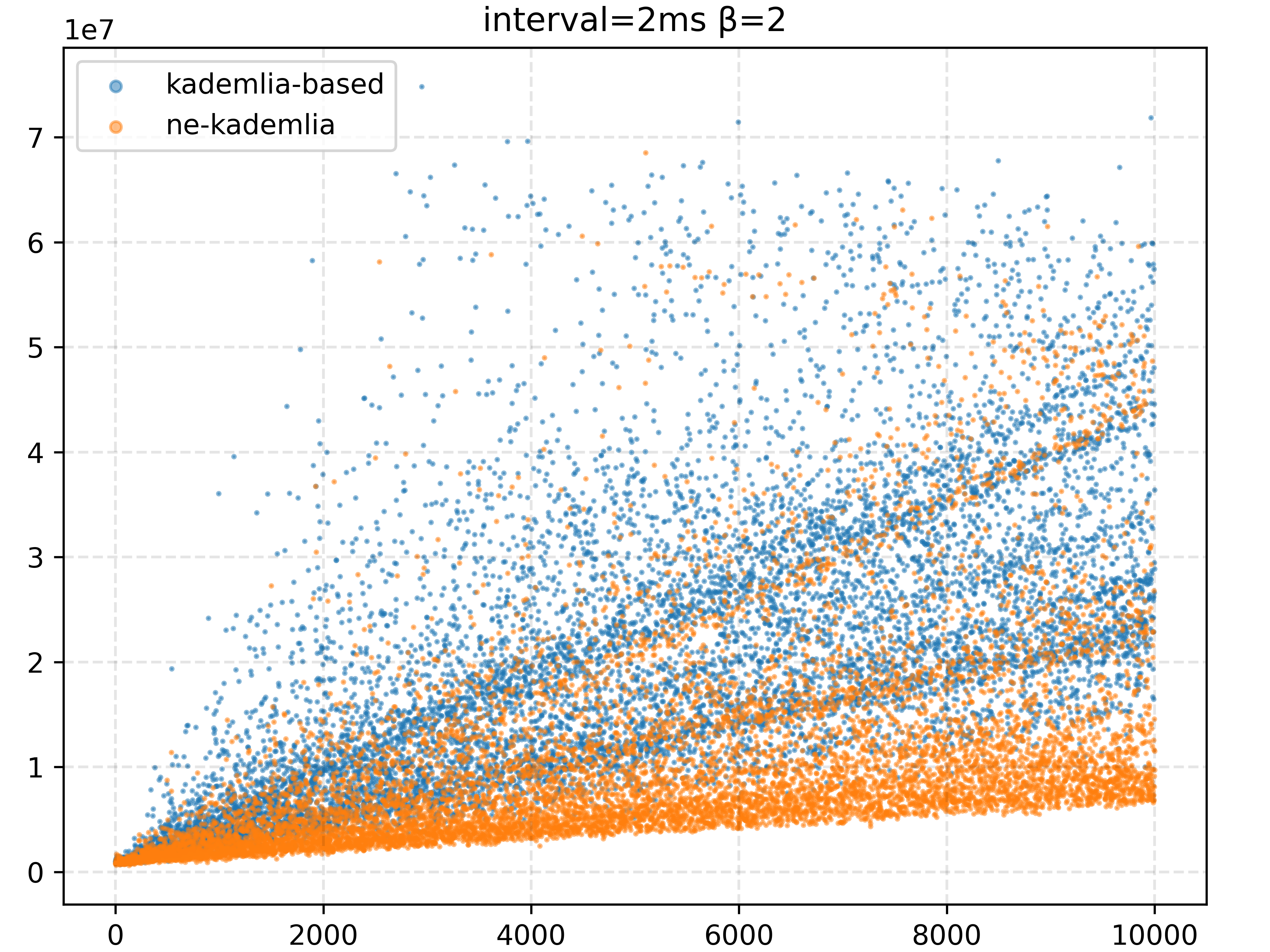}
    }
    \subfigure{
    \includegraphics[scale=0.4]{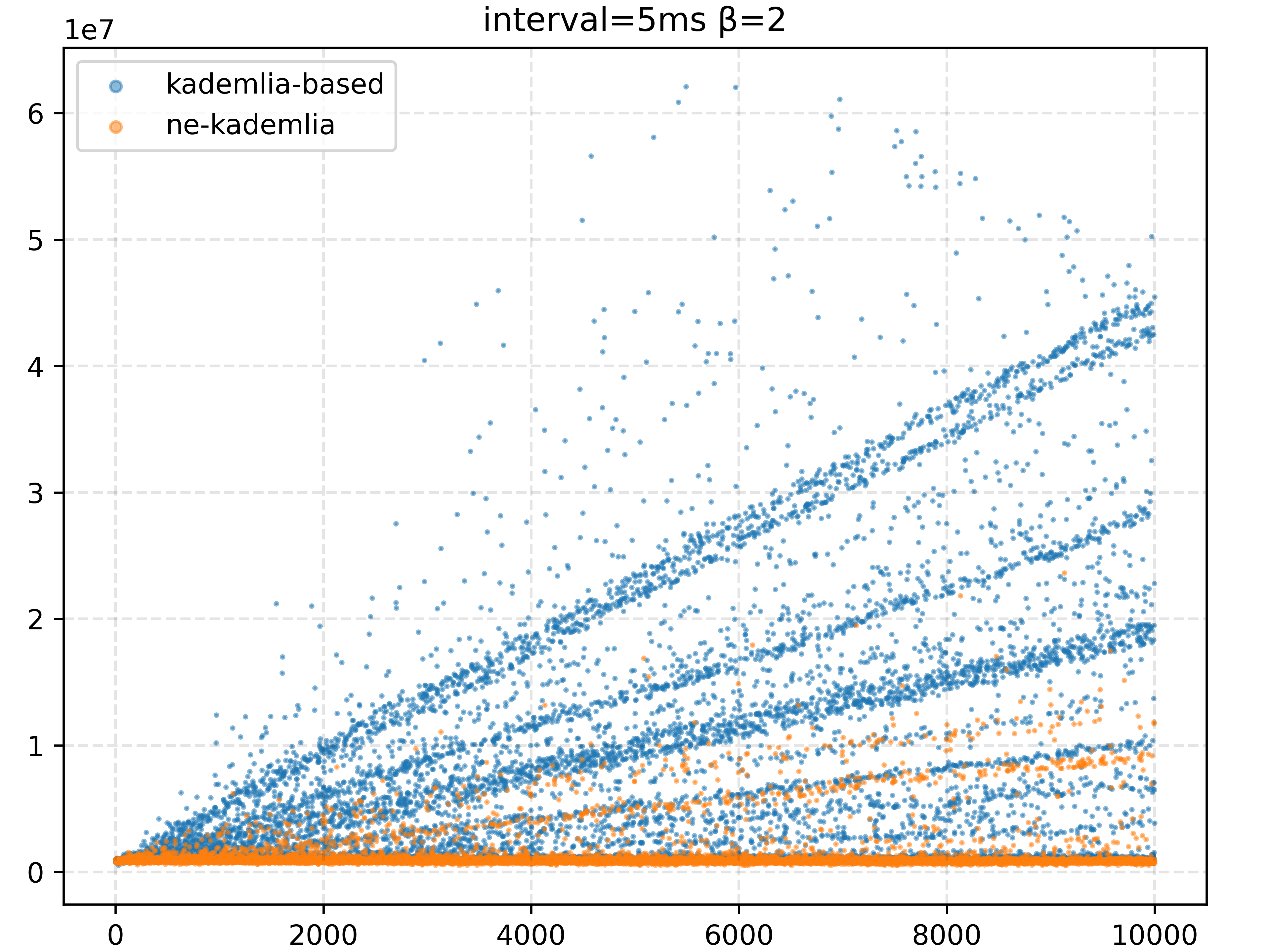}
    }
    \subfigure{
    \includegraphics[scale=0.4]{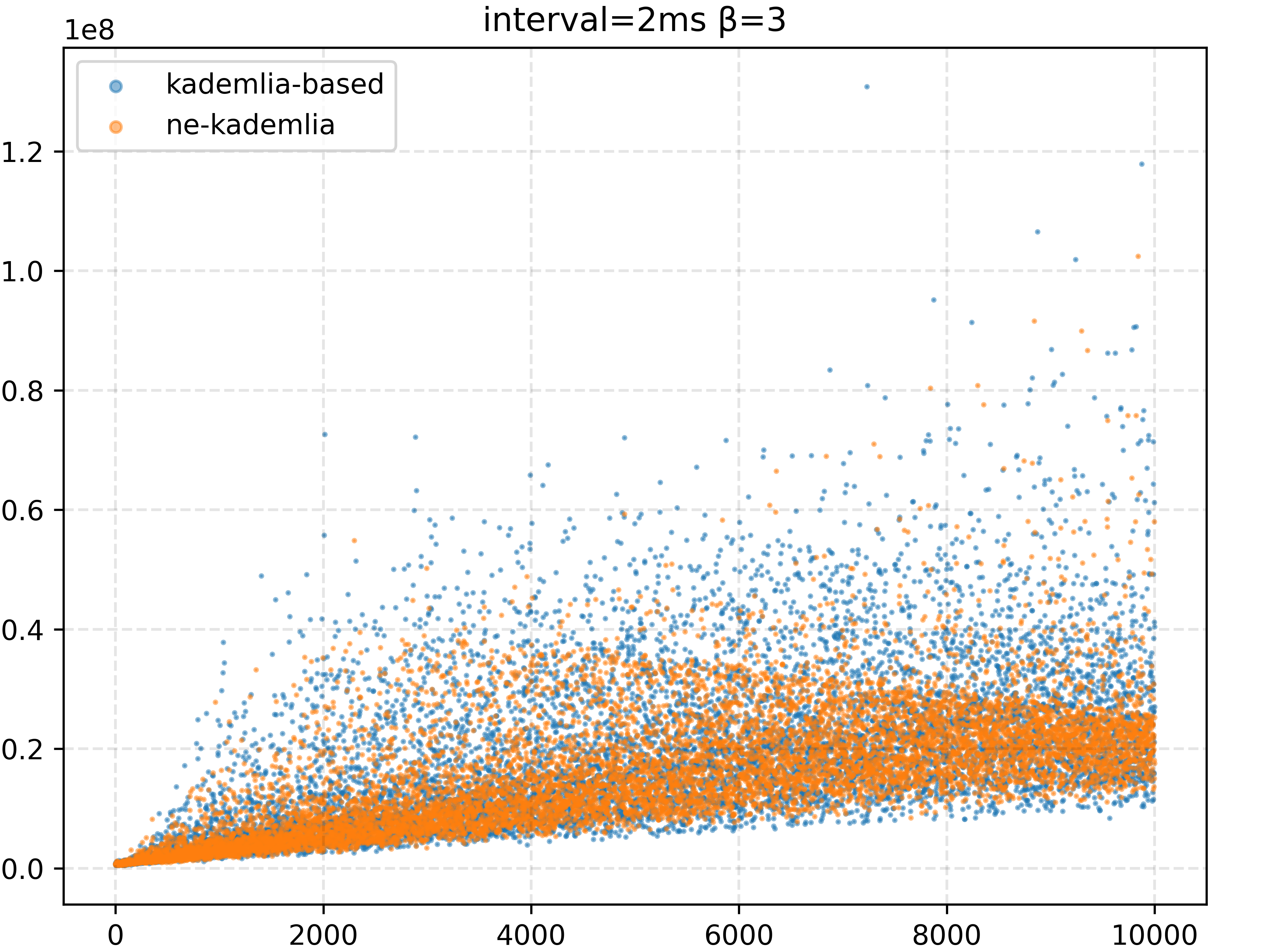}
    }
    \subfigure{
    \includegraphics[scale=0.4]{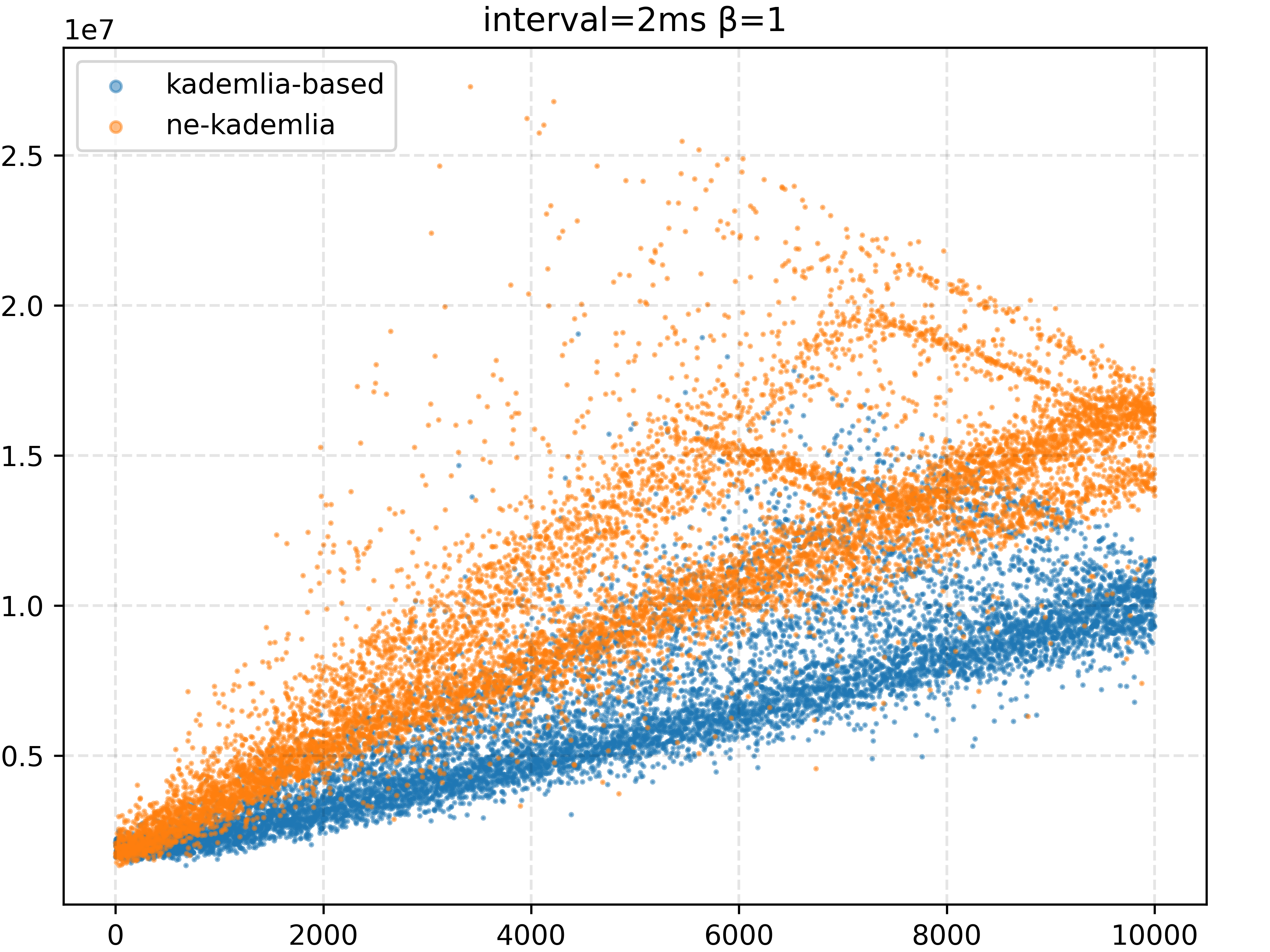}
    }
    \caption{In the figures, the blue dots represent the results of the broadcast algorithm based on Kademlia, while the orange dots represent the results of the Kademlia-based broadcast algorithm with NE mechanism. Each dot in the graph corresponds to one broadcast, and the horizontal coordinate of each dot represents its sequence number. Each algorithm has only one dot at each horizontal coordinate. The vertical coordinate represents the latency of the broadcast in the network, with the unit of $\mu$s, where 1e7 represents $10^7\mu$s, which is 10s, and 1e8 represents $10^8\mu$s, which is 100s. And $\beta$ represents the redundancy factor of the broadcast.}
\label{fig:result-latency}
\end{figure}

This part evaluates the latency of broadcast algorithms. As mentioned in section \ref{observations-and-design-principles}, the overall evaluation should be conducted by continuously performing many rounds of broadcast, rather than evaluating based on the latency of a single broadcast. During this process, all nodes remain online. The way to experiment is to initialize one broadcast to one node in the network in turn at a fixed interval. Each node initiated 10 broadcasts, for a total of 10,000 broadcast rounds per experiment.

\begin{figure}[htbp]
    \centering
    \includegraphics[scale=0.45]{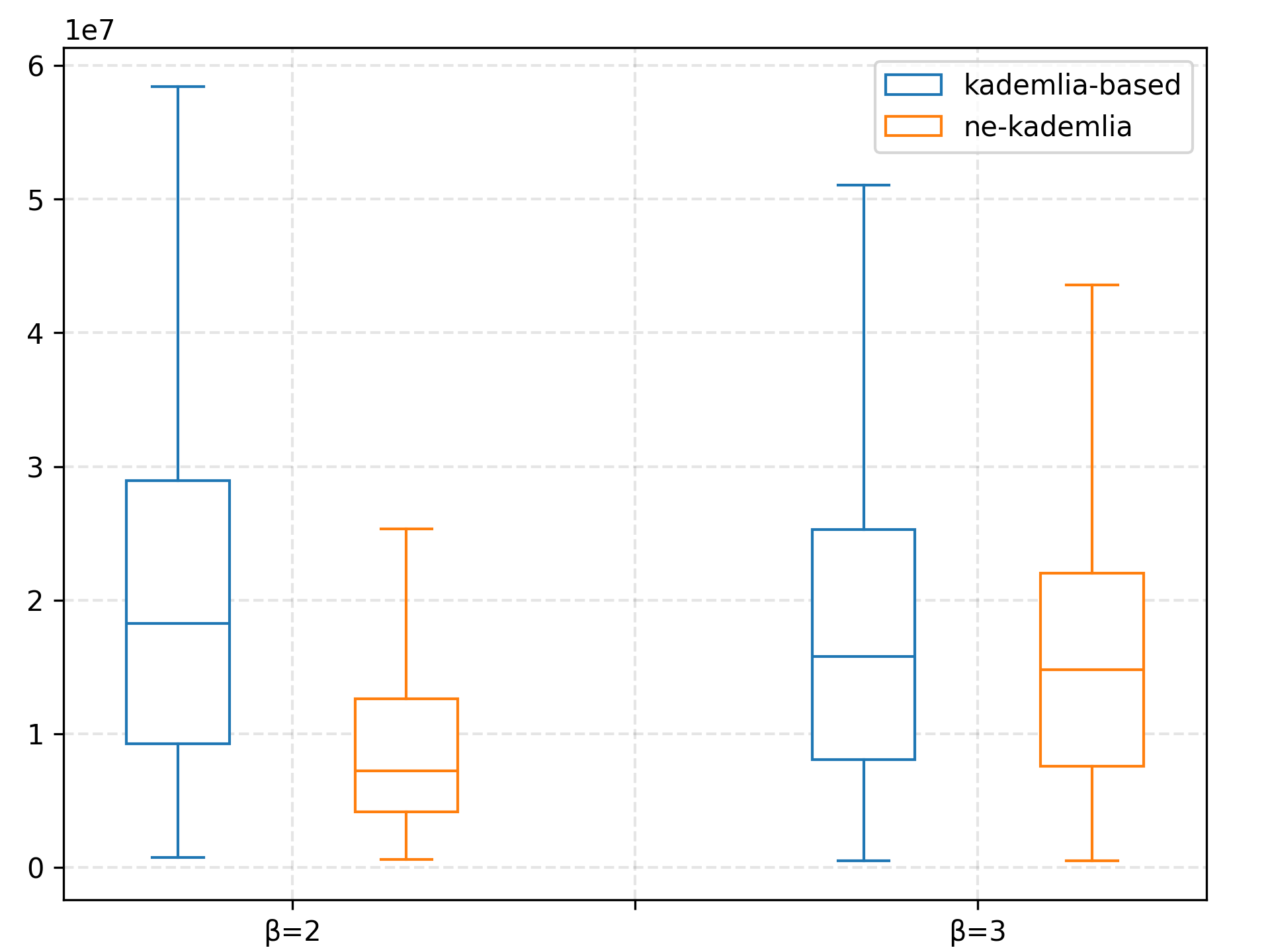}
    \caption{The delay comparison of the two algorithms corresponds to the first and third experimental results in Fig. \ref{fig:result-latency}.}
\label{fig:result-latency-comp}
\end{figure}

Fig. \ref{fig:result-latency} shows the results of four experiments with different parameters. The redundancy factor $\beta$ of the first two experiments is 2, and the broadcast initialization interval is 2ms and 5ms respectively. The broadcast initialization interval of the last two experiments is 2ms, and the redundancy factor $\beta$ is 3 and 1 respectively. Fig. \ref{fig:result-latency-comp} shows the delay comparison between the two algorithms when the interval is set to 2ms and $\beta$ is 2 and 3 respectively. It can be observed that when the redundancy factor $\beta$ is no less than 2 in the experiments, the NE mechanism shows a positive optimization effect on the latency. So, the NE mechanism enables more efficient propagation of broadcasts in the network. And the latency can be better optimized when the congestion level is within a certain range as the second experiment shows. When the redundancy factor $\beta$ is 1, the NE mechanism brings higher broadcast latency. This is because the source node selects only one relay node, and competition among multiple relay nodes is not formed. The relay node selected for the first time can obtain a higher score than other neighbor nodes, and thus it is more likely to be repeatedly selected, so that the node is unable to select a better relay node via the NE mechanism. In addition, because the NE mechanism needs to send extra redundant confirmation packets to the network, extra network resources are consumed.

The NE mechanism enables a node to select a neighbor node with a higher score in probability, and the score of the neighbor node derives from relaying a new message to the node earlier than other neighbor nodes, or propagating a broadcast message initialized by the node to the network with greater immediacy. Therefore, nodes with a higher score often indicates that the node and its successor nodes have a better network condition than the lower ones.

\subsection{Coverage Rate}
In this part, we evaluate the coverage rate of the broadcast algorithms under two scenarios: (1) when half of the nodes in the network are offline, and (2) when half of the nodes in the network refuse to relay messages.
\subsubsection{Coverage Rate in A Network with Half of Nodes Offline} \label{coverage-offline}
\begin{figure}[htbp]
    \centering
    \includegraphics[scale=0.30]{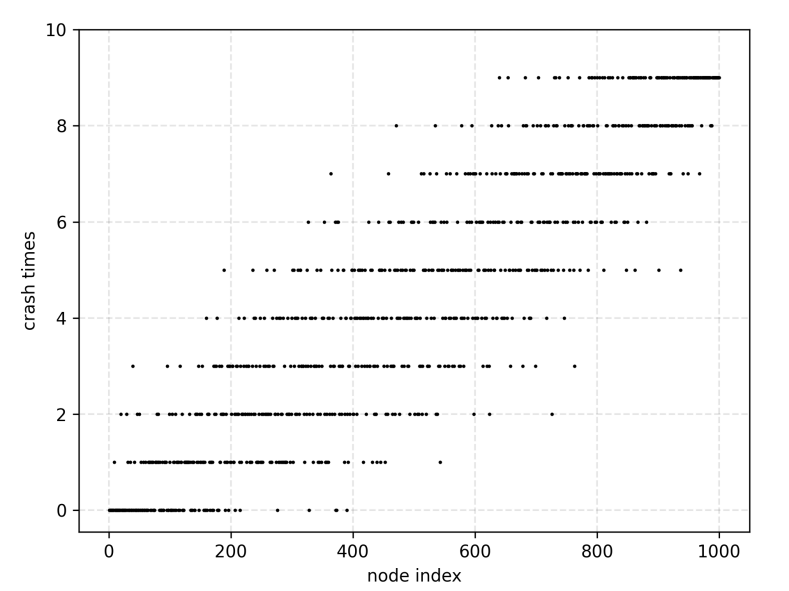}
    \caption{The node crash situation of the network in a single experiment. The horizontal axis represents the node's index, and the vertical axis represents the number of times the node crashed. The network underwent nine rounds of disturbance simulations in this experiment.}
\label{fig:crash-times}
\end{figure}

As mentioned in \cite{maymounkov2002kademlia}, the longer a node has been up, the more likely it is to remain up another hour. Therefore, we assume that different nodes can have different uptimes. In this experiment, the network has a size of 1000, and the probability of a node failure is linearly related to its serial number. That is, the node with serial number 1000 has a failure probability of 1, and the node with serial number 1 has a failure probability of 0.001. The offline or online status of nodes is maintained until the next network disturbance occurs. In this setting, nodes with lower serial numbers have longer uptimes, and about half of the nodes in the network are online at any given time. In Fig. \ref{fig:crash-times}, the number of crash times of 0 indicates that the node is online throughout the experiment, and the number of crash times of 9 indicates that the node is offline throughout the experiment.

The two algorithms were evaluated in the experiment under different redundancy factors $\beta$ of 1, 2, 3, and 4, as well as under the condition of a single network disturbance and with network disturbances occurring every 60 seconds. Each network disturbance takes about half of the 1000 nodes offline, clears their routing tables, and reestablishes neighbor relationships when they are set to online again to increase the effect of the disturbance on the network. After joining the network, nodes do not actively discover more new neighbors, but add non-neighbor nodes to their routing tables when receiving messages from them, and remove offline neighbor nodes from their routing tables. In each experiment, the network broadcasts 10,000 rounds with a fixed interval of 50ms. The coverage rate is calculated as follows:
\begin{equation}
    \frac{\sum_{i=1}^{M}N_{received_i}}{Round \cdot M \cdot N} \cdot 100\%
\end{equation}
where M represents the number of repeated experiments, which is 5, $N_{received_i}$ is the number of packets received by all online nodes in the i-th iteration of the repeated experiments, Round is the number of broadcasts for every node in each experiment, which is 10, and $N$ is 1000, the number of nodes in the network.

\begin{figure}[htbp]
    \centering
    \subfigure{
    \includegraphics[scale=0.45]{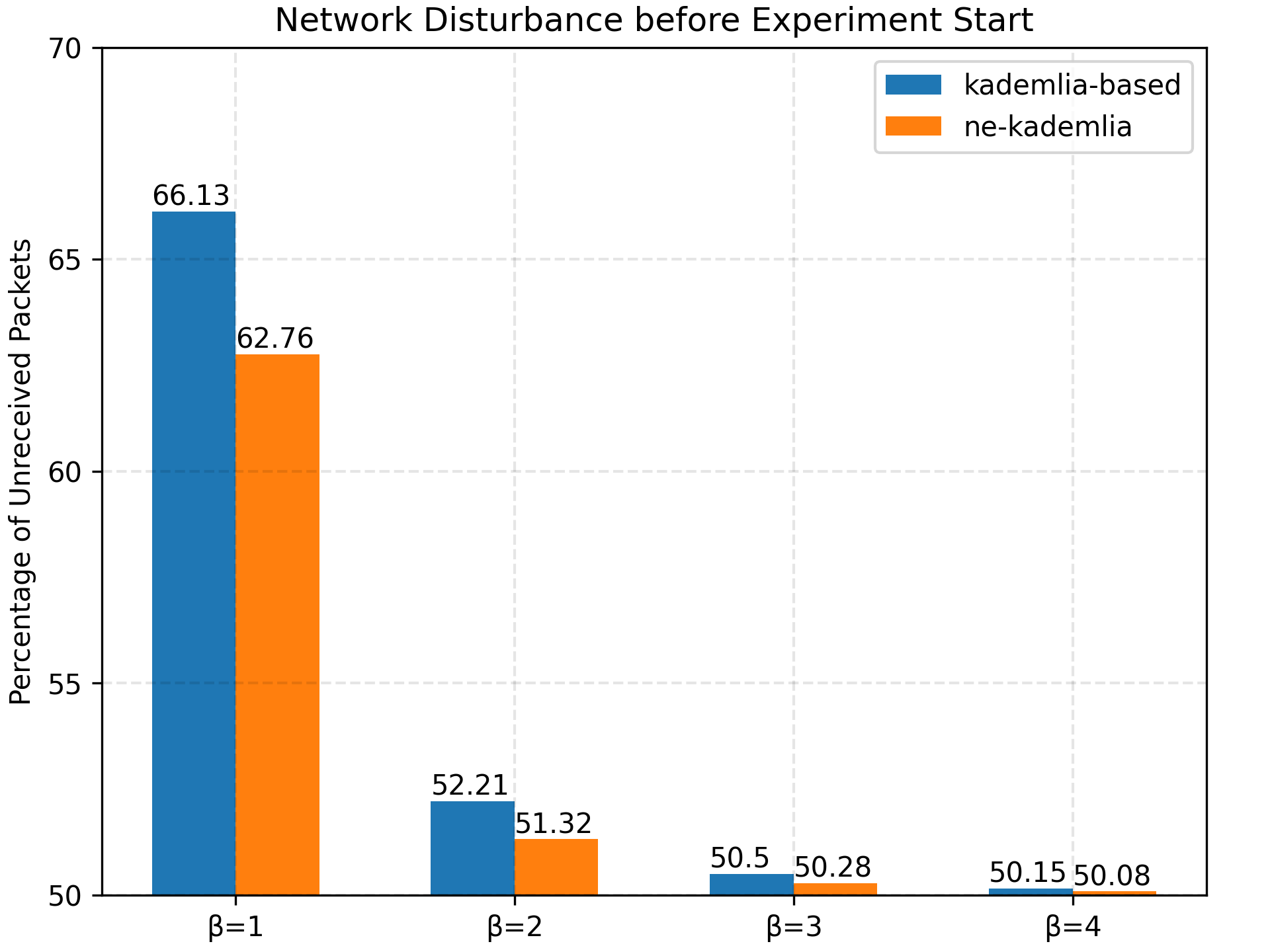}
    }
    \quad
    \subfigure{
    \includegraphics[scale=0.45]{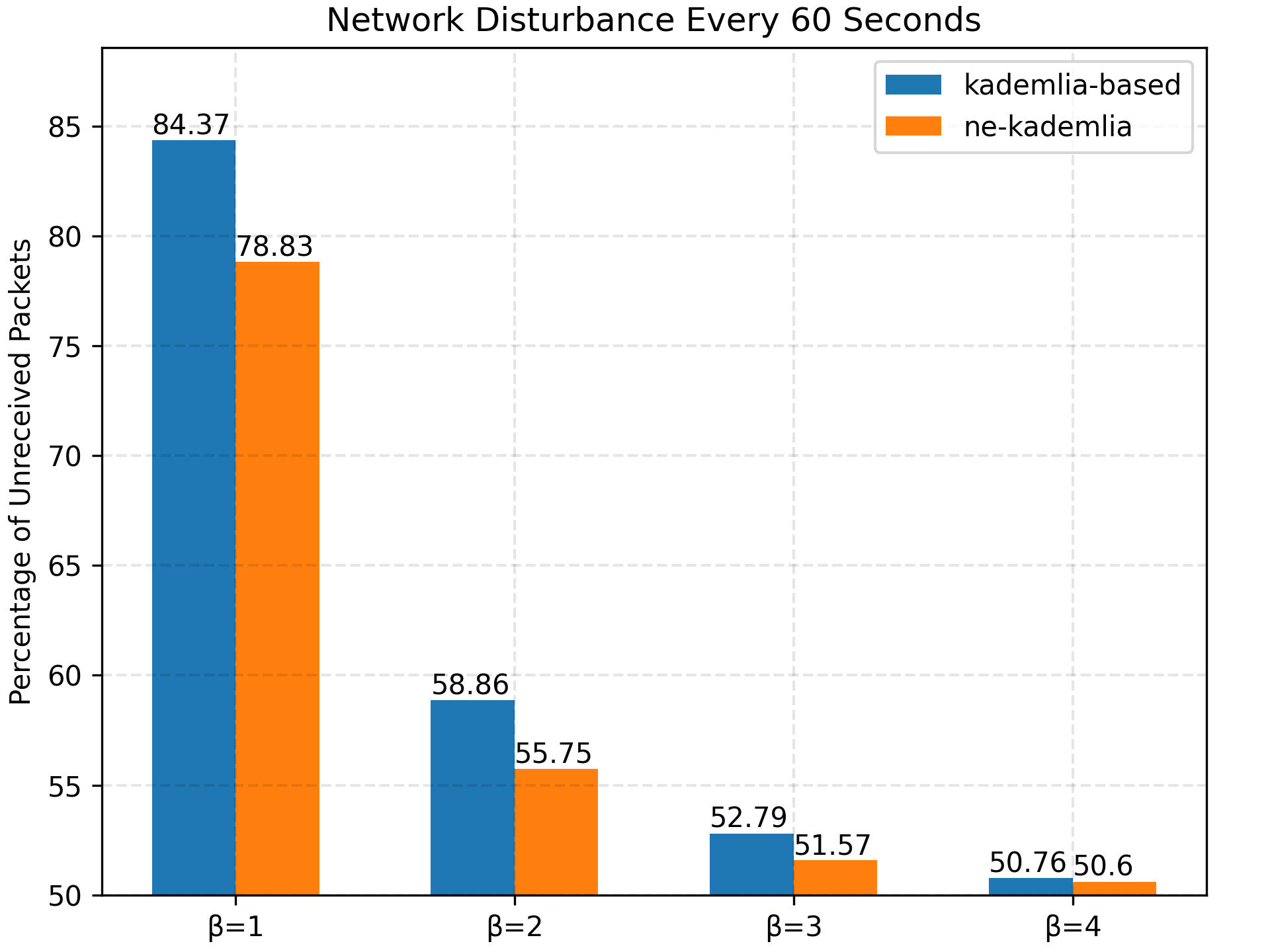}
    }
    \caption{The coverage rate of Kademlia-based broadcast algorithms with and without NE mechanism under different redundancy factors $\beta$ in a network where half of the nodes are offline. The first figure shows the experimental results of one network disturbance before the broadcast starts, and the right figure shows the experimental results of network disturbance once every 60s. The values on the vertical axis represent the proportion of packets that fail to be relayed to nodes in the network, so with lower values representing higher broadcast coverage rate.}
\label{fig:result-coverage-offline}
\end{figure}

The experimental results, shown in Fig. \ref{fig:result-coverage-offline}, indicate that the proportion of unreceived packet by the broadcast algorithm based on Kademlia is significantly reduced with the increase of the redundancy factor. It is also clear that the NE mechanism leads to higher coverage rate of the Kademlia-based broadcast algorithm in eight situations with different parameters. 

\subsubsection{Coverage Rate in A Network with Half of Nodes Refusing Packet Relaying}

The setup of this experiment is similar to that in Section \ref{coverage-offline}, with the difference being that all nodes in the network remain online, but half of them do not relay messages after receiving them. The coverage rate is calculated as follows:
\begin{equation}
    \frac{\sum_{i=1}^{M}N_{received_i}}{Round \cdot \sum_{i=i}^{M}N_{honest_i}} \cdot 100\%
\end{equation}
where $N_{received_i}$ is the number of packets received by all honest nodes in the i-th iteration of the repeated experiments, and $N_{honest_i}$ represents the number of honest nodes in the i-th iteration of the repeated experiments.

\begin{figure}[htbp]
    \centering
    \includegraphics[scale=0.45]{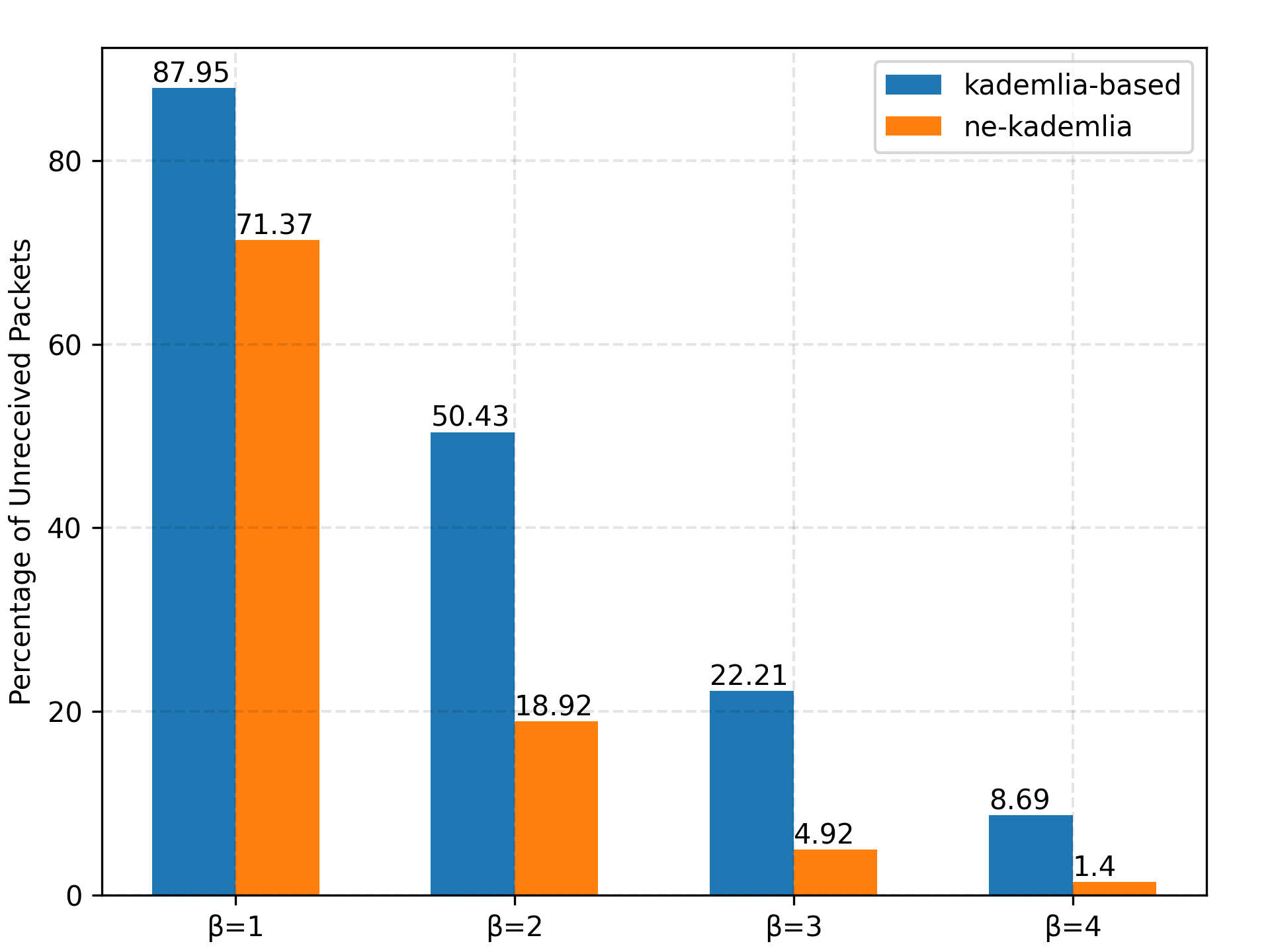}
    \caption{The coverage rate of the broadcast algorithm based on Kademlia with and without NE mechanism in a network where half of the nodes refuse to relay messages, when the redundancy factor $\beta$ is 1, 2, 3, 4, respectively.}
\label{fig:result-coverage-malicious}
\end{figure}

The experimental results are shown in Fig. \ref{fig:result-coverage-malicious}, which indicates that the NE mechanism enables nodes to more likely select better relay nodes. Since the Kademlia-based broadcast algorithm cannot identify whether neighbor nodes have relayed messages for them, the NE mechanism exhibits a greater improvement in broadcast coverage rate when nodes refuse to relay messages, compared to the scenario where nodes are offline.

\section{Conclusion} \label{section:conclusion}
In this paper, we propose the NE mechanism to dynamically evaluate the contribution of neighboring nodes to the interaction between a node and the network, enabling the selection of more reliable neighboring nodes for broadcasting messages. It facilitates faster and more reliable message propagation within the network, making the utilization of the DHT-based broadcast algorithm more practical for engineering applications.

According to the experimental results,  the NE mechanism has a positive effect on both the broadcast latency and the coverage rate of the Kademlia-based broadcast algorithm. In scenarios with a certain degree of network congestion, it can achieve noticeable improvements. And in situations where approximately half of the nodes are unreliable, it has been observed to reduce the number of unreceived packets by half in most of experiments.

\small
\bibliographystyle{IEEEtran}
\bibliography{refs}
\end{document}